\documentclass[showpacs,preprintnumbers,superscriptaddress,amsmath,amssymb]{revtex4}
\usepackage{epsf,psfig}
\usepackage{graphicx}
\usepackage{dcolumn}
\usepackage{bm}
\def\e{\epsilon}
\def\beq{\begin{eqnarray}}
\def\eeq{\end{eqnarray}}

\begin{document}
\title{Hydrodynamics of binary fluid phase segregation}
\author{Sorin Bastea}
\email{bastea2@llnl.gov}
\affiliation{Lawrence Livermore  National Laboratory, P.O. BOX 808, Livermore, CA 94550}
 \author{Raffaele Esposito}
 
\affiliation{Dipartimento di Matematica, Universit\`a di L'Aquila and
Centro di Ricerche
Linceo `Beniamino Segre', Roma, Italy}

\author{Joel L. Lebowitz}
\affiliation{Departments of Physics and  Mathematics, Rutgers
University, New Brunswick, NJ 08903}

\author{Rossana Marra}
\affiliation{Dipartimento di Fisica e Unit\` a INFM, Universit\`a di
Roma Tor Vergata, Roma, Italy}

\begin{abstract}
Starting with the Vlasov-Boltzmann equation for a binary fluid mixture, we
 derive an equation for the velocity field $\bm{u}$ when the system is
 segregated into two phases (at low temperatures) with a sharp interface
 between them.  $\bm{u}$ satisfies the incompressible Navier-Stokes
 equations together with a
 jump boundary condition for the pressure across the interface which, in
 turn, moves with a velocity given by the normal component of $\bm{u} $.
 Numerical simulations of the Vlasov-Boltzmann equations for shear flows
 parallel and perpendicular to the interface in a phase segregated mixture
support this analysis. We expect similar behavior in real fluid mixtures.
\end{abstract}


\maketitle

 When a fluid is quenched from a high temperature to a low temperature
 inside the miscibility gap it evolves from its initial homogeneous state
 which is now unstable into a final equilibrium state consisting of two
 coexisting phases separated by an interface.  The phase segregation
 process,
 involving hydrodynamical velocity fields ${\bf u}({\bf x},t)$, is more
 complicated than the corresponding diffusive processes
 in quenched binary alloys \cite{langer}.  In a seminal work
\cite{siggia} Siggia pointed
 out that in the late stages of phase segregation sharp interfaces develop
 between domains of the two phases and the coarsening process is mainly due
 to hydrodynamic flows driven by these elastic interfaces. Following
 Siggia there has been much work to investigate the time evolution of 
this coarsening process \cite{white,perrot,kendon}, and
 this is still going on \cite{solis,furukawa}.

There has been much less work on the derivation of a quantitative
 description of the velocity field involving the interface between two
 phases of macroscopic size, a problem of considerable interest in its own
 right \cite{anderson}.  To do this one should start from a microscopic
description
 of the transition region in which the
 composition varies smoothly on the microscopic scale and end up with
 boundary
 conditions on the ``membrane'' representing this transition layer on the
macroscopic scale (the
 sharp interface limit).  This is a difficult task for the case in which the
 two phases have widely different densities, viscosities, etc., since a
 reasonable microscopic description valid for both phases is hard to find.

To overcome this difficulty we consider here a situation in which both the
 equilibrium and kinetics are as simple as
 possible.  This is a binary fluid model introduced in \cite{bl} which
 provides a description of phase segregation kinetics at
 the mesoscopic level via the Vlasov-Boltzmann (VB) equations. These
 equations can be obtained from the microscopic dynamics (in
 a convincing but not rigorous manner) and in turn allow
 a derivation, in a suitable limit, of bulk hydrodynamical equations
\cite{belm1}.  The hydrodynamic
 equations
 we obtain for the bulk mixture are similar but not identical to the
heuristic
 macroscopic equations previously proposed for describing phase segregation
in an energy and
 momentum conserving system \cite {kawa}.  In the present paper we
consider the case in which the fluid
is already well
 segregated with sharp interfaces between the different phases.
 We derive hydrodynamic equations involving the interface motion,
 directly from the VB equations and carry out direct simulations for
 the time evolution of simple hydrodynamical flows, in the presence
 of such an interface. The resulting free boundary problem was considered
phenomenologically, in the context of interface oscillations, by Harrison
\cite{harrison}.
 We compare our results, with good agreement, with the behavior based on the
linearized version \cite{chandra} of the hydrodynamic equations, recently
considered for studying coarsening dynamics \cite{solis}. Both the
derivation and the numerical analysis are new and we believe
that the results have, as is the case of the Navier-Stokes
equations of a one component fluid, a universal validity even though their
rigorous derivation from microscopic models can only be carried out currently 
from the Boltzmann equation \cite{del}. In fact preliminary calculation
using the Enskog equation instead of the Boltzmann equation (more
appropriate for a dense fluid) leads to the same hydrodynamics.  

The model we consider is  a mixture of two species of hard spheres with
 diameter $d$ and unit mass
 labeled by $1,2$, with a long range
 positive potential $\ell^{-3} V(\ell^{-1} r)$ between particles of
different
 species; $\ell$ is the range of the interaction.
The latter causes phase
 segregation at temperatures lower than a critical temperature $T_c$,
into phases I and II:  I is rich in
 species 1 and II in species 2 \cite{ccelm}.  Let $n$ be the total particle
 density.  It was shown in \cite{belm1} that, when $nd^3 << 1$, while $
 \ell>> d$,
 this system is well described by coupled
 VB kinetic equations for the one-particle distributions
 $f_i(\bm{r}, \bm{v},\tau), i=1,2$
\begin{equation}
{\partial_{\tau} f_i}+ \bm{v}\cdot {\nabla}_{\bm r} f_i +\bm{F}_i\cdot
{\nabla}_{\bm{v}}
f_i=J[f_i,f_1+f_2]
\label{eq:one}
\end{equation}
where $\bm{F} _i(\bm{r},\tau)=-{\nabla}_{\bm r} V_i(\bm{r},\tau)$ with
$V_i$  the self-consistent
Vlasov potentials
$V_i(\bm{r},\tau)=\int _{\mathbb{R}^{3}}\ell^{-3}V(\ell^{-1}|
\bm{r}-\bm{r}'|)n_j(\bm{r}',\tau)d \bm{r}', i\ne j$, $n_j(\bm{r},\tau)=\int
_{\mathbb{R}^{3}}f_j(\bm{r}, \bm{v},\tau)d\bm{v}$,
and $J[f,g]$ is the Boltzmann collision operator for hard spheres
\cite{cip}.
We are interested in an equilibrium state at a fixed total density in which
the system is split into two regions, one consisting of phase I, and the
other of phase II.  These
regions are separated by a transition layer whose local form can be
obtained by
considering particular $1$-d stationary `solitonic' (actually kink)
solutions of (\ref{eq:one}): set $f_{i}=\rho_{i}(z)
M(\bm{v})
$ where
$M(\bm{v})$ is the Maxwellian with unit mass, zero mean velocity and
temperature $T$,
$M({\bf v}) = (2\pi k_B T)^{-3/2} \exp[-{\bm v} ^2/2k_BT]$. Moreover
$\rho_i(z)=\phi_i(z/ \ell) $ with $\phi_i(q)$
smooth functions satisfying the equations
\begin{equation}
k_{B}T\log  \phi_i(q)+ \int _{\mathbb{R}} d {q}'  \tilde V(|
{q}-{q}'|) \phi_j(q')=C_i
\label{three}
\end {equation}
where $\tilde V(q)= \int _{\mathbb{R} ^2} dw V(\sqrt{q^{2}+{w}^{2}})$ and
$C_i$ is a constant.  For $T < T_c$ Eqs. (\ref{three}) have
non-homogeneous solutions with asymptotic values $\rho^I_i$ and
$\rho^{II}_i$ as $q\to\pm\infty$, corresponding to the densities in
phases I and II respectively: by symmetry
$\rho_1^I = \rho^{II}_2$, $\rho^I_2 = \rho^{II}_2$. The solutions are
not explicit but their existence and properties are studied in
\cite{bl} and \cite{ccelm}. These are similar to those observed in realistic
mixtures, such as oil and water.
 
We are interested in studying hydrodynamical flows when the width
 of the interface, which is of order $\ell$, is small compared to the mean
free path
$\lambda$ which is in turn small
 compared to $L$, the characteristic length of the domains occupied by the
 pure phases.
 Setting
 $\bm{r} =\e^{-1} \bm{x}, \tau=\e^{-2}t$, where $\e =
\lambda/L\sim\ell/\lambda$ and
$\bm{x}$,
$t$ are the macroscopic position and time, we wish to
study the small
$\e$  behavior  of a solution
$f_i^\e(\bm{x}, \bm{v},t)=f_i(\e^{-1} \bm{x}, \bm{v},\e^{-2}t)$
of Eqs. (1), in the
incompressible regime, corresponding to low hydrodynamic velocity
\cite{del}. We assume far from the interface a solution of the form
\cite{del}
\begin{equation}
f_i^\e(\bm{x}, \bm{v},t)=\bar \rho_i(\bm x) M(\bm{v})+\sum_{s=1}^\infty
\e^s g_i^{s}(\bm{x} ,
\bm{v} ,t)
\end {equation}
where $\bar \rho_i({\bm x})$ is either $\rho^I_i$ or $\rho^{II}_i$
depending on
which region $\bm x$ is in.  Note that by symmetry
$\rho^I_1 + \rho^I_2 = \rho^{II}_1 + \rho^{II}_2 = \bar \rho$, i.e. the
total density
is the same in both phases but varies in the interface region. Putting 
$f^{\e}=f^{\e}_{1}+f^{\e}_{2}$, we
have for the
hydrodynamic velocity field
$$
\bar\rho\,^{-1}\int_{\mathbb{R}^{3}} \bm{v} f^\e(\bm{x}, \bm{v},t)
d\bm{v}=
\e
\  \bm{u}(\bm{x},t) +O(\e^2),
$$
so that the Mach number is of order $\e$ (incompressible regime).  Far from
the interface the temperature $T$ and the densities  are constants,
coinciding for $T < T_c$ with the asymptotic values of the solitonic
solutions.

To understand what happens near the interface consider the situation in
which the interface is flat. Then $f_i(\bm{r}, \bm{v},\tau)= \rho_i(z)
M(\bm{v})$ with $\rho_i(z)=\phi_i(z/ \ell) $ a stationary solution of
the Vlasov-Boltzmann equations. If the interface is not flat this is not
exactly true, but because of the tendency of the solitonic profile to
keep its form and just translate, one expects that the solution is
locally close to the solitonic profiles multiplied by the Maxwellian. On the
other hand, the force due to the surface tension acts on the particles of
the fluid which start to move with some average velocity $\bm{u}$ and 
the surface
moves accordingly, so that its normal velocity is at any point
$\bm{u}\cdot
\bm{N}$, $\bm{N} $ being the normal to the surface. To implement this
picture we write the solution near the interface as
\begin{equation}
f_i^\e(z,\tilde{\bm x},\bm{v},t)=\rho_i(z) M(
\bm{v})+\sum_{s=1}^\infty\e^{s}\tilde
g_i^{s}(z,\tilde{\bm x}, \bm{v},t).
\label{interface}
\end{equation}
The notation is as follows: given a point $\bm{x} $ we call
$z=\e^{-1}d(\bm{x},\Gamma_t)$,  where $d (\bm{x},\Gamma_t)$ is the
distance of $\bm{x} $ from the interface $\Gamma_t$, and $\tilde{\bm{x}} $
the component of
$\bm{x} $ tangential to the interface;  $\rho_i(z)=\phi_i(z/ \ell)$ is the
solitonic solution centered on $z = 0$.  Since we are so close to the
interface we can assume that locally it looks almost flat, so that the
solution near the interface has a weak dependence on $\tilde {\bm x}$.
Therefore the profile
interpolating between the values of the densities in the bulk on the two
sides of the interface should be well approximated by the 1-d solitonic
profile $\rho_{i}(z)$, the lowest order term in Eq. (\ref{interface}).

To obtain a solution of Eqs. (1) we have to put these expressions in the
 equations (after space-time rescaling) and match the two expansions.
This can be done in a consistent way and it is possible to compute the 
terms in the series (the long computation will be reported elsewhere), 
but the question of the convergence of the series is open (we expect 
they are asymptotic). The result is that in the limit $\e\to 0$ the velocity field
 $\bm{u}(\bm{x},t)$ is divergence free and solves the incompressible
 Navier-Stokes (INS) equation
\begin{equation}
{\partial_{t} \bm{u}}+ (\bm{u}\cdot \bm{\nabla}) \bm{u} +{\bm \nabla}
p=\nu \Delta {\bm u}
\label{NS}
\end {equation}
with the kinematic  viscosity $\nu$
obtained from the Boltzmann equation as in \cite{cip}.  Moreover, 
$\bm{u} $ is
continuous across the interface $\Gamma_t$ whose normal velocity is
given by
\begin{equation}
v_{\Gamma_t}(\bm{x})= \bm{u}(\bm{x},t)\cdot \bm{N}(\bm{x},t)
\end {equation}
while the pressure is discontinuous at the surface and satisfies
Laplace's law
\begin{equation}
(p_+-p_-)=\sigma K,
\label{motion}
\end {equation}
Here ($p_{-}$)  $p_+$  is the pressure on  the side of  $\Gamma_{t}$
(not) containing  the normal $\bm N$; $K$ is the
mean curvature of $\Gamma _{t} $ and $\sigma$ is the surface tension
given in terms of solitonic profiles as
\begin{equation}
\sigma=\frac{1}{2}\int (q'-q) \sum_{i\ne  j=1,2 }\frac{d
\phi_i(q)}{d q}\tilde V(q-q')\phi_j(q')dqdq'
\end {equation}

The equation for  $\bm{u}$ is a complicated initial boundary value
problem with free boundary $\Gamma_t$.
The solutions of INS in the two domains separated by $\Gamma_t$ are
coupled by the jump
condition for the pressure. Since the solution is not explicit, we
compare our numerical
results with the linearized version of this problem, studied in \cite{chandra}.

To obtain numerical solutions of the VB equations we use the method
introduced in \cite{bl} to simulate the VB
dynamics at the particle level, based on coupling the direct simulation
Monte Carlo algorithm for the short-range interaction with the
grid-weighting method for the long-range repulsion. The results were obtained using
about $5\times
10^{6}$ particles, in a cube ($L\times L\times L$) with periodic boundary
conditions. The unit of length is the mean free path
$\lambda=(2^{\frac{1}{2}}\pi \bar\rho d^2)^{-1}$, and the unit of time is
the mean-free time between collisions
$\tau=\lambda/c$, $c=(2k_BT)^{\frac{1}{2}}$.
For the repulsive inter-species potential we use, as in \cite{bl},
$V(q) = \alpha\pi^{-\frac{3}{2}} \exp (-q^{2})$ (note that the mean-field
critical temperature is given by $k_BT_c=\bar\rho\alpha/2$), with a
range of
interaction $\ell=0.4\lambda$. All quenches were performed at total
particle density $\bar\rho$, with $\bar\rho d^{3}\approx 0.01$.
We focus on the behavior of a simple initial  flow
perturbation, $u_z(0)=u_0 \cos(k y)$, with $k=2\pi n_k/L$, $n_k = 1, 2,
3, ...$, outside and inside the coexistence region; for all simulations
we set $u_0=0.1c$.   For a homogeneous system this shear wave
perturbation leads to a time dependent flow
$u_z(t)=u(t)\cos(k y)$, where $u(t)=u(0)\exp(-\nu k^2 t)$,
$\nu=\eta/\bar\rho$ and
$\eta$ is the viscosity.

We first study the behavior of the above initial perturbation in the
homogeneous region
of the phase diagram at $T/T_c=1.5$. The initial velocity
profile decays exponentially with very high accuracy and we
extract a kinematic
viscosity $\nu$ which agrees with the Boltzmann gas viscosity \cite{cip}.
We then turn to the study of the system in the presence of interfaces,
at two
temperatures, $T_1/T_c=0.51$ and $T_2/T_c=0.33$. For each of these
temperatures the
initial configuration contains two interfaces (due to periodic boundary
conditions),
situated in the $\{xy\}$ plane a distance $L/2$ apart, that separate
domains of the two equilibrium
phases for the particular temperature. The static structure of these
interfaces has been described
in \cite{bl}.  The thickness  of the interface is of order $\ell$.

For an initial flow perturbation that is parallel with the interfaces,
${\bf u}(0) = u_x(0){\hat {\bf x}}=u_0 \cos(k y){\hat {\bf x}}$, where
$\hat{\bf x}$ is the unit vector in the ${x}$-direction,  the behavior
of the velocity profile is virtually identical  with the one in the
homogeneous region. The situation is however very different
if the initial perturbation is perpendicular to the interfaces, i.e.
${\bf u}(0) = u_z(0)\hat {\bf z} =u_0 \cos(k y){\hat {\bf z}}$.
While $u_z(t)$ is still very well represented by $u(t)\cos(k y)$, $u(t)$
no longer decays exponentially.
To fully characterize this situation we also look at the position of the
interface.
The order parameter $\varphi=(\rho_1-\rho_2)/\rho$ is calculated at time 
as a function of $z$ and $y$. We find that $\varphi(z, y, t)$ is very well fitted by the
profile $\varphi_0\tanh[(z-z_0(y,t))/\xi]$
\cite{bl}, and identify $z_0(y, t)$ with the position of the interface.
Furthermore, $z_0(y, t)$ is well
represented by $z_0(y,t)=A(t)\cos(\kappa y)$.
The velocity amplitude $u(t)$ and the interface amplitude $A(t)$ for the
temperature $T_1$ and $n_k=1$
are shown in Fig. 1, along with fits with the functional forms expected
for an overdamped harmonic oscillator:
$A(t)=u_0\exp(-\gamma t)\sinh(\omega t)/\omega$,
$u(t)=u_0\exp(-\gamma t)[-\gamma \sinh(\omega t)/\omega + \cosh(\omega
t)]$.

We find that these forms can
reasonably reproduce the results up to the highest $n_k$ simulated,
$n_k=9$. In Fig. 2 we show the amplitudes
of the interface oscillations $A(t)$ up to $n_k=3$, along with the corresponding
best fits, for temperature $T_1$(top) and $T_2$(bottom).
We note that the $n_k=1$ behavior at $T_2$ is qualitatively different
from the other cases, showing damped oscillations.
The functional forms that describe this result are:
$A(t)=u_0\exp(-\gamma t)\sin(\omega t)/\omega$,
$u(t)=u_0\exp(-\gamma t)[-\gamma \sin(\omega t)/\omega + \cos(\omega
t)]$.

These simulation results can be compared with those obtained from the
linear perturbation analysis of a thin, initially flat
interface with surface tension $\sigma$ separating fluids of matching
density $\bar\rho$ and
viscosity $\eta=\nu\bar\rho$. This type of analysis has been carried out in
\cite{chandra} in the full space.
The idea is to look for flow solutions $u_{x,y,z}(t)\propto
\hbox{\rm e}^{\mu t}\cos(k_x x + k_y y)$
following a small velocity perturbation of wave number $k$.  This
yields a $u_z$ decaying exponentially  as a function of the distance from
the interface. We carried out a similar analysis matching the
periodic boundary and initial conditions in which $u_z$ is
independent of $z$ away from the interfaces.  Under these conditions 
and taking into account the
discontinuity of the normal stresses due to surface tension, we can
write, following
\cite{chandra} (see e.g. Chapter X, Eq. 28)
\begin{equation}
\rho{\partial_{t} u_z}=\eta\Delta u_z +
\Sigma_s[\sigma({\partial_{x}^2}+
{\partial_{y}^2})\delta z_s]\delta(z - z_s)
\end{equation}
where $z_s$ is the position of the interface $s$, $\delta z_{s}$ is the 
displacement from the initial position and $d\delta
z_s/dt=u_{zs}$ is
the velocity of the interface. However, we regard the above equation
integrated over the torus, along $z$,
as the more fundamental one if $u_z$ is $z$ independent. This yields
\begin{equation}
\rho{\partial_{t} u_z}=\eta\Delta u_z +
{L^{-1}}\Sigma_s\sigma({\partial_{x}^2}+
{\partial_{y}^2})\delta z_s
\end{equation}
Using $u_z=u_{zs}$ and assuming $u_z(t)\propto \hbox{\rm e}^{\mu
t}\cos(k y)$, which is
consistent with the initial conditions, we get
$\mu^{2}\rho+\mu k^2\eta+ {L}^{-1}{2\sigma k^2}=0$.
This equation is characteristic of a damped oscillator.
Introducing the length scale $L_s=\nu^{2}\rho/\sigma$ and time scale
$T_s=\nu^{3}\rho^{2}/\sigma^{2}$, this becomes
\begin{equation}
\mu^{2}_*+\mu_*k^{2}_*+ {L_*} ^{-1}{2k^{2}_*}=0
\end{equation}
where $\mu_*=\mu T_s$, $k_*=kL_s$ and $L_*=L/L_s$. Using $k=2\pi n_k/L$
we find that the above equation
predicts damped oscillations for $n_k<n^{0}_k=
(2L_*)^{\frac{1}{2}}/\pi$
and overdamped behavior
(no oscillations) otherwise. For our binary fluid model the surface
tension can be  related
to the interface profile (see Eq. 8) and easily estimated \cite{bpl}, and
therefore we can calculate
$n^{0}_k$. We find that for $T_1/T_c=0.55$, $n^{0}_k=0.83$, so that for
all $n_k$
the oscillations should be overdamped, in agreement with the simulation
results. For
$T_2/T_c=0.33$, $n^0_k=1.39$, so the $n_k=1$ case should exhibit damped
oscillations,
while the others should be overdamped; this is again in agreement with
the simulations.
 We also calculate the coefficients $\gamma$ and $\omega$
that describe the behavior of the velocity profile and compare with the
simulation results, see Fig. 3.
 We find  good agreement  for small $k$ ($n_k$) and increasing
deviations as $k$ increases; the relative behavior of $\gamma$ and
$\omega$ also appears to be qualitatively different
from the simulations for large $k$.
This is as expected, since for larger $k$, i.e.
smaller  wavelengths, we are far from the hydrodynamic
 regime. Our hydrodynamic picture is valid if $k\lambda=2\pi n_k\e$,
$\ell/\lambda$  and $u/c$ are of order
$\e=\frac{\lambda}{ L}$, which is in our simulations   $ {0.028}$.
Better agreement  is found in simulations with
$\ell/\lambda=0.2$ (and $\epsilon=.056$), see Fig. 3, which indicates
that  the accuracy of  the equations we derive increases when the
interface becomes sharper.

The work S.B. was performed under the auspices of the U. S.
Department of  Energy by University of California Lawrence Livermore
National Laboratory under  Contract No. W-7405-Eng-48.  The work of J. L.
was supported by NSF Grant DMR 98-13268, AFOSR Grant AF 49620-01-1-0154,
DIMACS and its supporting agencies, the NSF under contract STC-91-19999 and
the N.J. Commission on Science and Technology, and NATO Grant
PST.CLG.976552. R.M. partially supported by
MIUR, Cofin-2000 and INFM.

\newpage
                  
\begin{figure}
\includegraphics{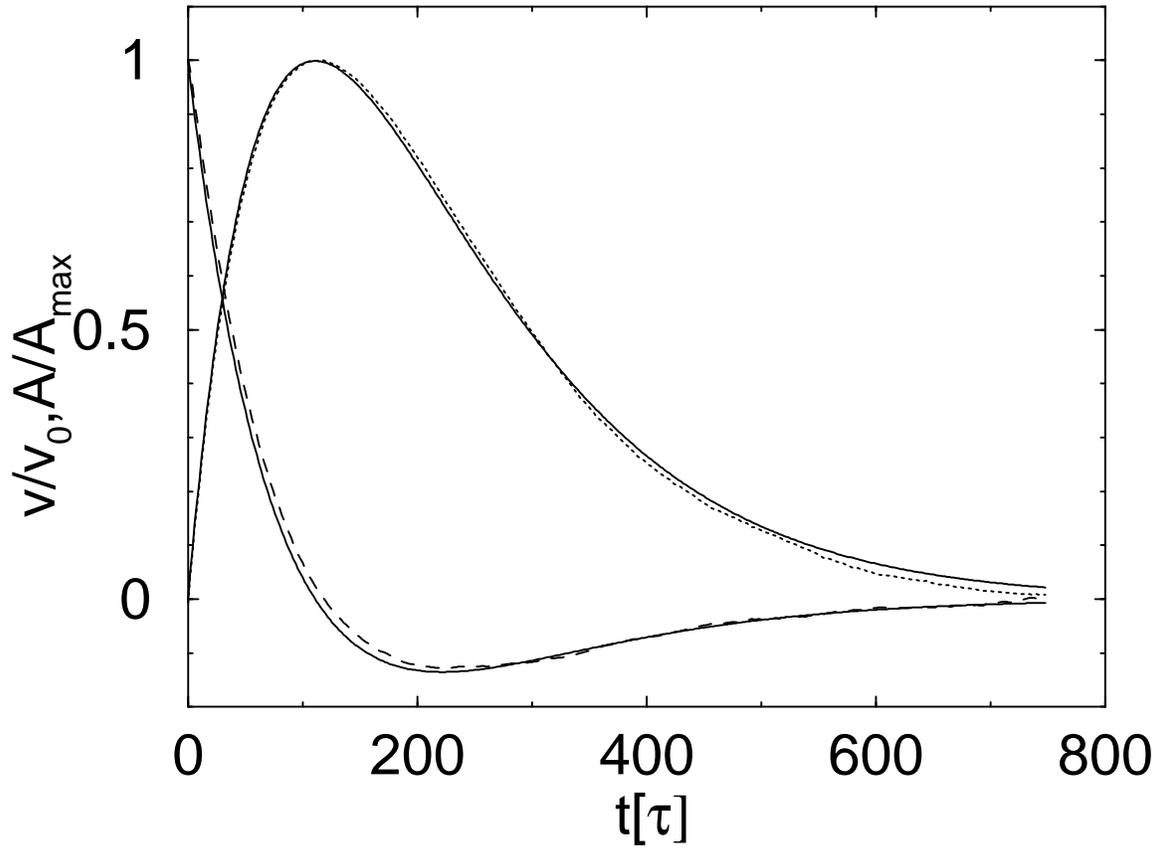}
\vspace{0.1in}
\caption{Interface amplitude $A(t)$ (dotted line) and velocity amplitude
$u(t)$ (dashed line) for $T_1/T_c=0.51$, $n_k=1$ and corresponding fits 
(see text) (solid lines).}
\end{figure}
%
\begin{figure}
\includegraphics{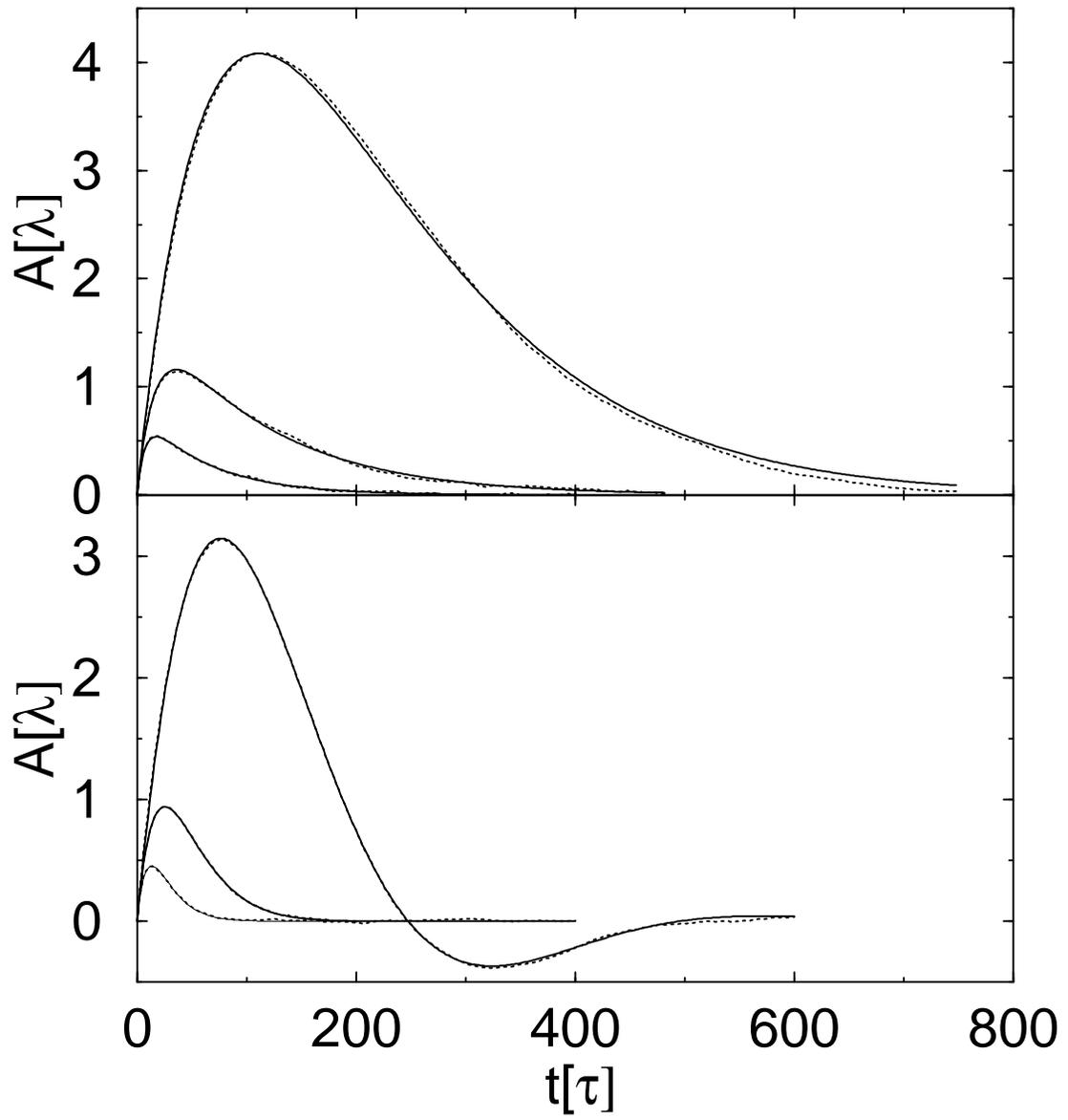}
\vspace{0.1in}
\caption{Interface amplitudes $A(t)$ for $T_1/T_c=0.51$ (top) and $T_2/T_c=0.33$ (bottom), 
$n_k=1,2,3$ (top to bottom) (dotted lines) and corresponding fits (see text) (solid lines).}
\end{figure}
%
\begin{figure}
\includegraphics{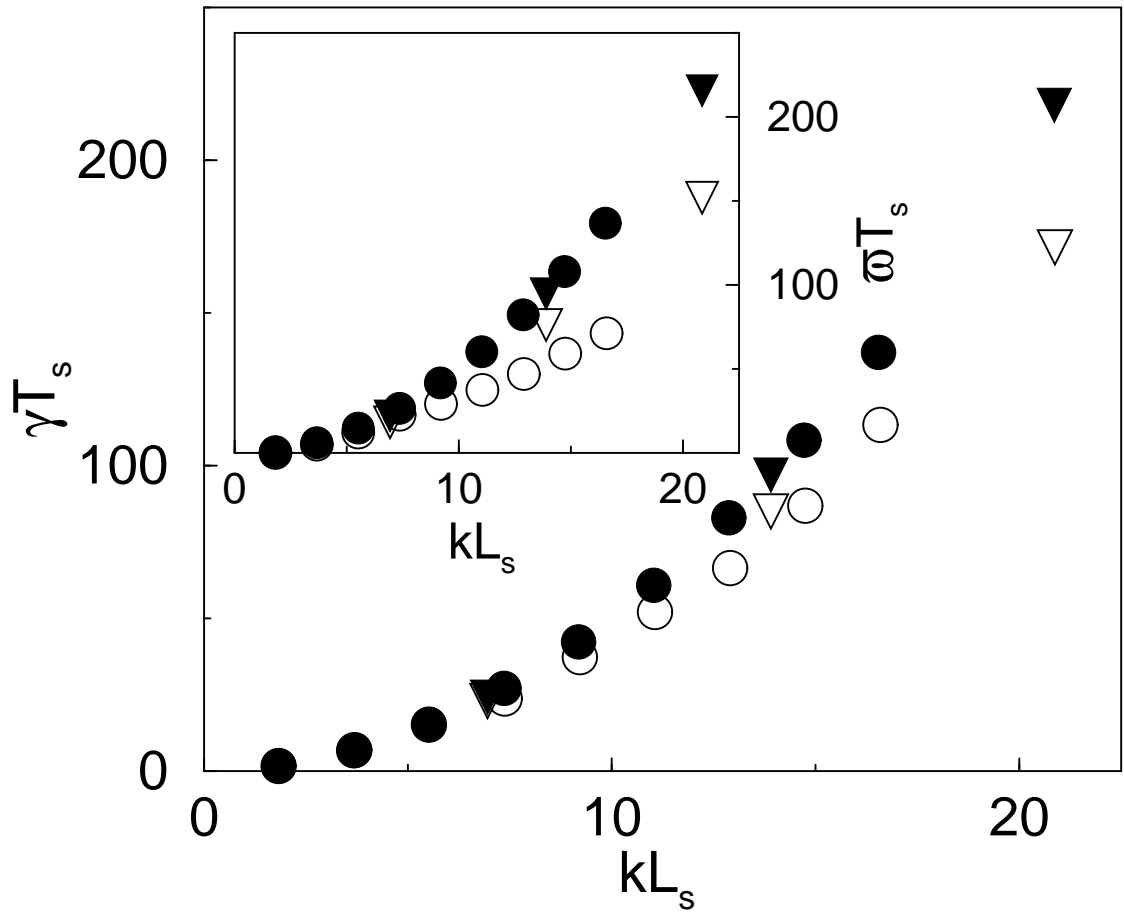}
\vspace{0.1in}
\caption{Coefficients $\gamma$ and $\omega$(inset): from simulations (open symbols) at 
$T_1$ with $\ell=0.4\lambda$ (circles) and $\ell=0.2\lambda$ (triangles); 
oscillator equation (see text) (filled symbols).}
\end{figure}
                      
\end{document}